\documentclass[fleqn,usenatbib]{mnras}
\usepackage{mathptmx}
\usepackage[varvw]{newtxmath}  
\usepackage[T1]{fontenc}
\usepackage{ae,aecompl}
\usepackage{times}
\usepackage{amsmath}
\usepackage{upgreek}
\usepackage{xcolor}
\usepackage{graphicx}
\usepackage{pdflscape}
\usepackage{multicol} 
\usepackage{amsmath,bm}
\usepackage{caption}

\begin{document}

\title[The most massive star clusters in molecular clouds]{The most massive star clusters in molecular clouds: Insights from the integrated cloud-wide initial mass function (ICIMF) theory}

\author[J. W. Zhou]{
J. W. Zhou \thanks{E-mail: jwzhou@mpifr-bonn.mpg.de}$^{1}$
Pavel Kroupa \thanks{E-mail: pkroupa@uni-bonn.de}$^{2,3}$
Sami Dib \thanks{E-mail: dib@mpia.de}$^{4}$
\\
$^{1}$Max-Planck-Institut f\"{u}r Radioastronomie, Auf dem H\"{u}gel 69, 53121 Bonn, Germany\\
$^{2}$
Helmholtz-Institut f{\"u}r Strahlen- und Kernphysik (HISKP), Universität Bonn, Nussallee 14–16, 53115 Bonn, Germany \\
$^{3}$
Charles University in Prague, Faculty of Mathematics and Physics, Astronomical Institute, V Hole{\v s}ovi{\v c}k{\'a}ch 2, CZ-180 00 Praha 8, Czech Republic \\
$^{4}$Max Planck Institute f\"{u}r Astronomie, K\"{o}nigstuhl 17, 69117 Heidelberg, Germany
}

\date{Accepted XXX. Received YYY; in original form ZZZ}
\pubyear{2024}
\maketitle

\begin{abstract}
The combination of the high-resolution ALMA, JWST and HST observations provides unprecedented insights into the connection between individual molecular clouds and their internal stellar populations in nearby galaxies. The molecular clouds in five nearby galaxies were identified based on the integrated intensity maps of CO (2$-$1) emission from ALMA observations. We used the JWST 21 $\mu$m data to estimate the star formation rate (SFR) surface density of the clouds and calculate the masses of the embedded stellar populations in the clouds. After matching the star cluster and stellar association catalogs derived from the HST observations with the identified molecular clouds, we found clear correlations between the physical parameters of molecular clouds and their internal stellar populations. Based on the masses of the total stellar populations and their corresponding clouds, we obtained a typical value of the cloud-scale star formation efficiency (SFE), $\approx$1.4\%.  The mass of the most massive cluster ($M_{\rm cluster, max}$) in a cloud is positively proportional to the mass ($M_{\rm cloud}$), the column density, the SFR sand the SFR surface density of the cloud. The observed $M_{\rm cluster, max}$-$M_{\rm cloud}$ relation can be interpreted theoretically on the basis of the integrated cloud-wide IMF (ICIMF) theory, which provides a quantitative framework for understanding the correlations between molecular clouds and their internal stellar populations.  
\end{abstract}



\maketitle

\section{Introduction}\label{sec:intro}

\begin{figure*}
\centering
\includegraphics[width=1\textwidth]{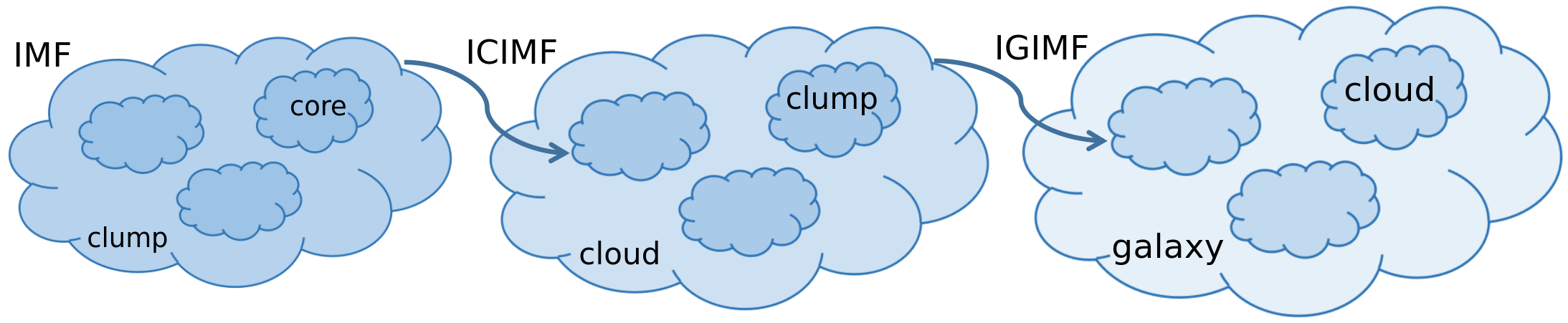}
\caption{Correspondence between hierarchical gas structures (star-forming regions) and hierarchically integrated IMFs. "ICIMF" and "IGIMF" are the integrated cloud-wide IMF and the integrated galaxy-wide IMF, respectively.}
\label{layer}
\end{figure*}

\begin{figure*}
\centering
\includegraphics[width=0.97\textwidth]{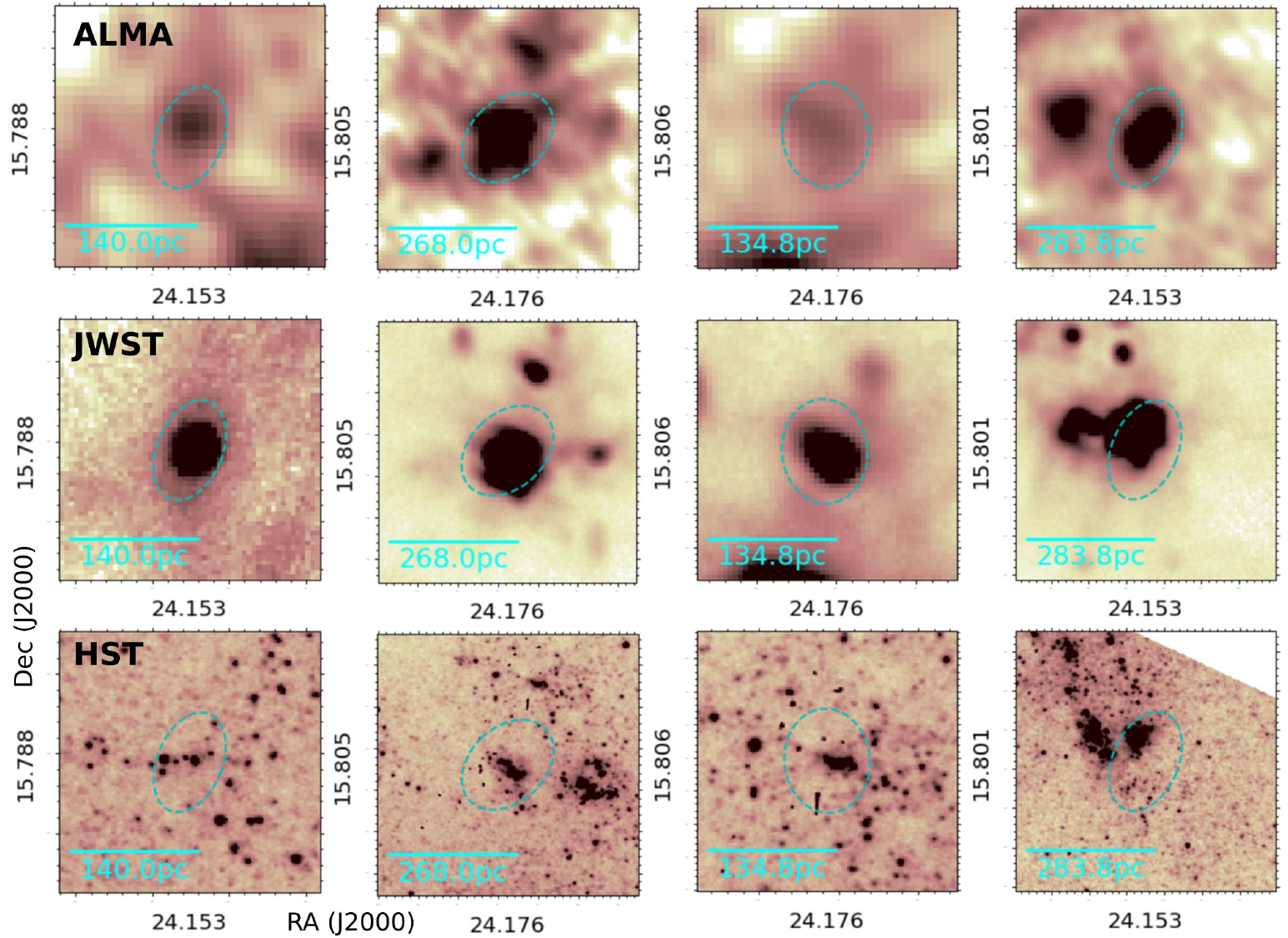}
\caption{Some examples from \citet{Zhou2024MNRAS.tmp.2058Z} are used to demonstrate the spatial distribution of molecular clouds in ALMA observation (the first row) and stellar populations in JWST (the second row) and HST (the third row) observations. Dashed cyan ellipses represent the effective ellipses of the molecular clouds.}
\label{case}
\end{figure*}

Galaxies act as stellar factories in the universe, producing stars through the gravitational collapse of the densest regions within molecular clouds in the interstellar medium (ISM). Molecular clouds are widely distributed within galaxies and serve as localized sites for star formation. Each molecular cloud typically contains multiple clumps, which are the local sites of star formation within the cloud and the precursors of embedded star clusters \citep{Dib2012ApJ-758,Miville2017-834,Rosolowsky2021-502,Urquhart2022-510,Zhou2023-676}.
The hierarchical structure of molecular gas in a galaxy determines the spatial distribution of stellar populations \citep{Gouliermis2015MNRAS-452, Elmegreen2014-787,Grasha2017-840,Chevance2023-534}. 
Young clusters with different ages constitute a cluster population within the parent molecular cloud. The evolution of the molecular cloud is closely related to the formation and evolution of the internal star clusters. The star clusters become fully exposed after the molecular cloud is dispersed by feedback from the star clusters \citep{Kim2016-819,Kruijssen2019-569,Chevance2022-509,Watkins2023-676,Zhou2024-682-173}.

It has been suggested that most if not all observed stars originate from embedded clusters \citep{Kroupa1995a-277, Kroupa1995b-277, Lada2003-41,Kroupa2005-576,Megeath2016-151, Dinnbier2022-660}. Based on this viewpoint,
the integrated galaxy-wide initial mass function (IGIMF) theory has been developed to calculate galaxy-wide IMFs, starting with the IMF that is observationally constrained at the star cluster scale \citep{Kroupa2003-598,
Kroupa2013-115,Yan2017-607,Jerabkova2018-620,Dib2022-666,Haslbauer2024arXiv}.
The IMFs assembled in star-forming regions combine to create the IMF of the entire galaxy. This process involves first generating a series of embedded star clusters from the embedded cluster mass function (ECMF), then producing a series of stars for each embedded cluster from the IMF, and finally summing the stars from all clusters to construct the galaxy-wide IMF. 
Star clusters first form in molecular clouds. Molecular clouds, serving as intermediaries between star clusters and the galaxy, should also be incorporated into the IGIMF theory. Molecular clouds and their internal star clusters could be quantitatively characterized using a localized IGIMF theory, which can be called the integrated cloud-wide initial mass function (ICIMF) theory. Fig.\ref{layer} illustrates the correspondence between hierarchical gas structures (star-forming regions) and hierarchically integrated IMFs, providing a depiction of their relationship.


With high-resolution observations across various wavelengths, we are now able to break down star formation into individual molecular clouds in nearby galaxies \citep{Leroy2021-257,Leroy2021-255,Lee2022-258,Emsellem2022-659,Lee2023-944}. High-resolution CO imaging data from ALMA (Atacama Large Millimeter/submillimeter Array) observations enable systematic studies of molecular clouds across nearby galaxies. The JWST (James Webb Space Telescope) and HST (Hubble Space Telescope) observations can reveal embedded clusters, young star clusters, and stellar associations within molecular clouds \citep{Gouliermis2017-468,Whitmore2021-506,Thilker2022-509,Larson2023-523,Levy2024arXiv}. The combination of these high-resolution observations provides an unprecedented look at the connection between individual clouds and the stellar populations \citep{Turner2022-516}. Some examples of multi-band observations are shown in Fig.\ref{case}.
In this work, we first look for correlations between the physical parameters of molecular clouds and their internal stellar populations, and then use the ICIMF theory to interpret the observed correlations. 

\section{Data and catalogs}

\subsection{Star cluster and stellar association catalogs} \label{catalog}

The catalogs for compact clusters and multi-scale associations across five galaxies (NGC 1433, NGC 1566, NGC 1559, NGC 3351 and NGC 3627) are being made available as the DR3 release of the PHANGS-HST survey. A detailed description of these catalogs can be found in \citet{Lee2022-258,Whitmore2021-506,Thilker2022-509,Larson2023-523}.
Based on the classification technique, there are two kinds of compact cluster catalogs, i.e. human classified and machine learning (ML) classified. The ML classified technique greatly increases the number of identified star clusters within each galaxy, allowing for a much better statistical analysis. Therefore, we primarily use the machine learning-classified catalogs in this work, while also employing the human-classified cluster catalogs to verify the results. 
According to NUV–U–B–V–I photometry of the star clusters, the publicly available spectral energy distribution (SED) fitting code CIGALE was used to characterize their physical properties, such as their age, mass, and reddening \citep{Turner2021-502}.
For the stellar association catalogs,
as described in \citet{Larson2023-523},
they firstly selected point sources from the DOLPHOT source catalogs \citep{Thilker2022-509} to serve as ‘tracer’ stars,
using both the HST NUV-band (F275W) and V-band (F555W). 
The positions of ‘tracer’ stars are used to create star density maps on which they used the watershed algorithm to identify stellar associations. All light from within the bounds of an association is used in the photometry for that structure. The measured five-band NUV, U, V, B, I photometry gives the total flux of the structure. Using the SED fitting \citep{Turner2021-502}, they estimated the ages, masses, and reddening of the stellar associations according to the total flux.
The associations are identified at three physical scale levels (16 pc, 32 pc and 64 pc) for all five galaxies. 
The NUV-band selection primarily identifies the youngest stellar populations, while the V-band traces both young and old stars, enabling the detection of stellar associations across a wider range of ages. As described in \citet{Larson2023-523}, the V-band selection identifies more associations at all scale levels. Most of the associations are identified through V-band selection, with 75-85 percent of the young associations found in the NUV-band also overlapping with the larger set of associations detected in the V-band. 

In this work, we exclude clusters and associations whose estimated masses are less than their mass estimation errors, as their masses are not well determined. Considering that the typical survival timescale of GMCs is 10$-$30 Myr \citep{Chevance2021,Kim2022-516}, an age cut of 20 Myr is imposed to eliminate chance superpositions between older stellar populations and molecular clouds.

\subsection{CO (2-1) cube and 21 $\mu$m map} 

\begin{table*}
\centering
\caption{Resolutions of the datasets. NGC 1559 is not observed by JWST currently. "D" and "$i$" are the distances and the inclination angles of the galaxies, respectively.}
\begin{tabular}{cccccc}
\hline
	&D (Mpc)	&$i$ (deg) &angular (")&linear (pc)&linear (pc)\\
 	&		&	&ALMA&	ALMA&	JWST	\\
NGC 1433	&	18.6&28.6	&	1.10	&99.1	 &58.6 \\
NGC 1559	&	19.4&65.4	&	1.25	&117.5	 &-- \\
NGC 1566	&	17.7&29.5	&	1.25	&107.6	 &55.7 \\
NGC 3351	&	10.0&45.1	&	1.46	&70.7	 &31.5 \\
NGC 3627	&	11.3&57.3	&	1.63	&89.2	 &35.6 \\
\hline
\label{data}
\end{tabular}
\end{table*}

For the five galaxies,
we used the combined 12m+7m+TP PHANGS-ALMA CO (2$-$1) data cubes to investigate gas kinematics and dynamics, which have a spectral resolution of 2.5 km s$^{-1}$. We also use the JWST 21 $\mu$m (F2100W) maps from the PHANGS-JWST survey. The full width at half maximum (FWHM) of the original F2100W point spread function (PSF) is $\approx$0.65" \citep{Williams2024-273}.
The angular resolutions or linear resolutions of these datasets are listed in Table.\ref{data}.
Overviews of the PHANGS-ALMA and PHANGS-JWST surveys' science goals, sample selection, observation strategy, and data products are described in \citet{Leroy2021-257,Leroy2021-255,Lee2023-944}. 
All the data are available on the PHANGS team website \footnote{\url{https://sites.google.com/view/phangs/home}}. 

\section{Results and discussion}\label{res}

\subsection{Identification of molecular clouds}\label{dendro}

\begin{table}
\centering
\caption{Leaf structures in each galaxy, see Sec.\ref{comp} for more details.}
\begin{tabular}{ccccc}
\hline
	&	leaves	&	fitted	&	type1	&	type3	\\
NGC 1433	&	187	&	178	&	175	&	3	\\
NGC 1559	&	262	&	253	&	246	&	7	\\
NGC 1566	&	588	&	559	&	540	&	19	\\
NGC 3351	&	271	&	264	&	263	&	1	\\
NGC 3627	&	471	&	458	&	414	&	44	\\
total	&	1779	&	1712	&	1638	&	74	\\
\hline
\label{type}
\end{tabular}
\end{table}

\begin{figure*}
\centering
\includegraphics[width=1\textwidth]{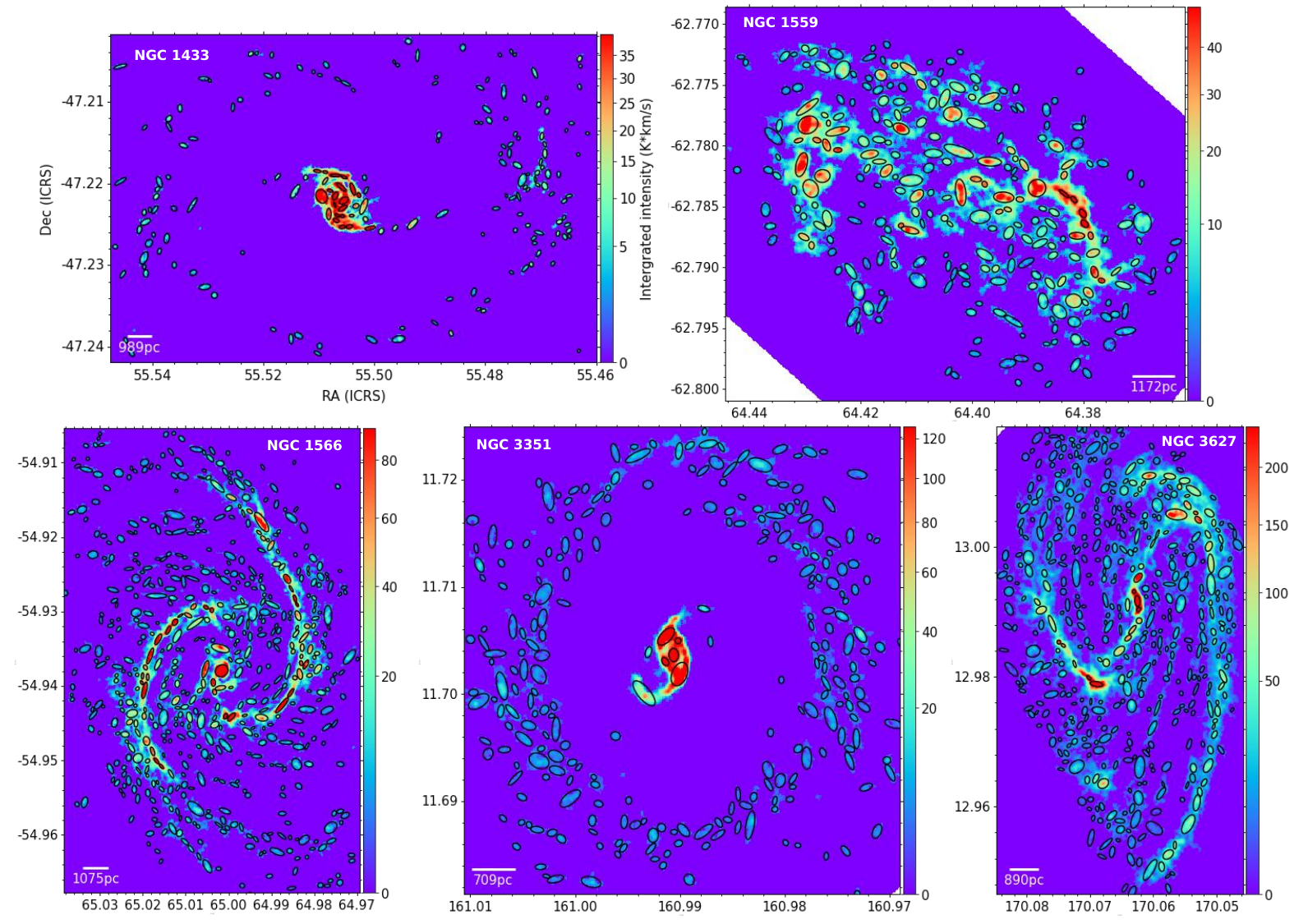}
\caption{Background is the intergrated intensity map of CO (2$-$1) emission. Black ellipses represent leaf structures identified by the dendrogram algorithm. The scalebar shows the scale of 10 beam sizes.}
\label{co}
\end{figure*}

For the five galaxies considered in this work, we conducted a direct identification of hierarchical (sub-)structures based on the 2D intensity maps. As described in \citet{Rosolowsky2008-679}, the dendrogram algorithm decomposes density or intensity data into hierarchical structures called leaves, branches, and trunks. 
Using the {\it astrodendro} package \footnote{\url{https://dendrograms.readthedocs.io/en/stable/index.html}},
there are three major input parameters for the dendrogram algorithm: {\it min\_value} for the minimum value to be considered in the dataset, {\it min\_delta} for a leaf that can be considered as an independent entity, and {\it min\_npix} for the minimum area of a structure.
For the CO (2$-$1) data cube, 
there are two types of Moment 0 maps (strictly masked and broadly masked) in the data product of the PHANGS-ALMA survey \footnote{Details of the masking strategy and completeness statistics are presented in the PHANGS pipeline paper \citep{Leroy2021-255}.}. The strictly masked maps only include emissions that are identified as signals with high confidence in the data cube, which might filter out the relatively faint structures. The broadly masked maps offer superior completeness and cover larger areas compared to the strictly masked maps. However, due to the inclusion of more regions with faint emissions or areas close to bright emissions, they tend to be noisier and may contain false positives. 
In order to ensure the reliability of the identified structures and because we are only interested in local dense structures, we select the strictly masked Moment 0 maps to identify structures.
Since all the retained structures on the strictly masked Moment 0 maps are reliable, we only require the smallest area of the identified structure be larger than 1 beam area. We do not set additional parameters in the algorithm to minimize the dependence of the identification on parameter settings. Finally, the numbers of leaf structures in the five galaxies are listed in Table.\ref{type}. In Fig.\ref{co},  
the CO structures identified by the dendrogram algorithm exhibit a strong correspondence with the background intensity maps.
For a structure with an area $A$ and a total integrated intensity $I_{\rm CO}$, the mass of the structure can be calculated by
\begin{equation}
M_{\rm cloud} = \alpha^{2-1}_{\rm CO} \times I_{\rm CO} \times A,
\label{mass}
\end{equation}
where $\alpha^{2-1}_{\rm CO} \approx 6.7~\rm M_{\odot} \rm pc^{-2} (\rm K~km~s^{-1})^{-1}$ \citep{Leroy2022-927}.

For the 21 $\mu$m emission, apart from the {\it min\_npix}= 1 beam area, we also take 
the values of {\it min\_value}= 3*$\sigma_{\rm rms}$, {\it min\_delta} = 3*$\sigma_{\rm rms}$, where $\sigma_{\rm rms}$ is the background intensity.
We first retained as many structures as possible using these lower standards, then eliminated the diffuse structures, as done in \citet{Zhou2024MNRAS.tmp.2058Z}.

The algorithm characterizes the morphology of each structure by approximating it as an ellipse. Within the dendrogram, the root mean square (rms) sizes (second moments) of the intensity distribution along the two spatial dimensions define the long and short axes of the ellipse, denoted as $a$ and $b$. As described in \citet{Zhou2024-682}, the initial ellipse with $a$ and $b$ is smaller, so a multiplication factor of two is applied to appropriately enlarge the ellipse.
Then the effective physical radius of an ellipse is $R_{\rm eff}$ =$\sqrt{2a \times 2b}*D$, where $D$ is the distance of the galaxy.

\subsubsection{Velocity component}\label{comp}

For the identified molecular structures, we extracted the average spectrum of each structure to investigate its velocity components and gas kinematics. 
Large-scale velocity gradients from the galaxy's rotation contribute to the non-thermal velocity dispersion. To address this, before extracting the average spectra, we removed the bulk motion due to galaxy rotation by creating gas dynamical models using the Kinematic Molecular Simulation (KinMS) package \citep{Davis2013-429}.  
According to the averaged spectra, 67 structures with poor line profiles were eliminated.
Then, following the procedure described in \citet{Zhou2024-682-128}, we fitted the averaged spectra of 1712 leaf structures individually using the fully automated Gaussian decomposer \texttt{GAUSSPY+} algorithm \citep{Lindner2015-149, Riener2019-628}. 
As shown in Table.\ref{type}, $\approx$96\% structures have a single velocity component, i.e. type1 (following the nomenclature introduced in \citet{Zhou2024-682-128}). 
For some structures with more than one velocity component (i.e. type3), as discussed in \citet{Zhou2024-682-173}, we treat type3 structures as independent structures like type1 structures, but with more complex gas motions. 

\subsection{Correlations between molecular clouds and their internal stellar populations}

\subsubsection{Stellar associations}\label{s-assoc}

\begin{figure}
\centering
\includegraphics[width=0.47\textwidth]{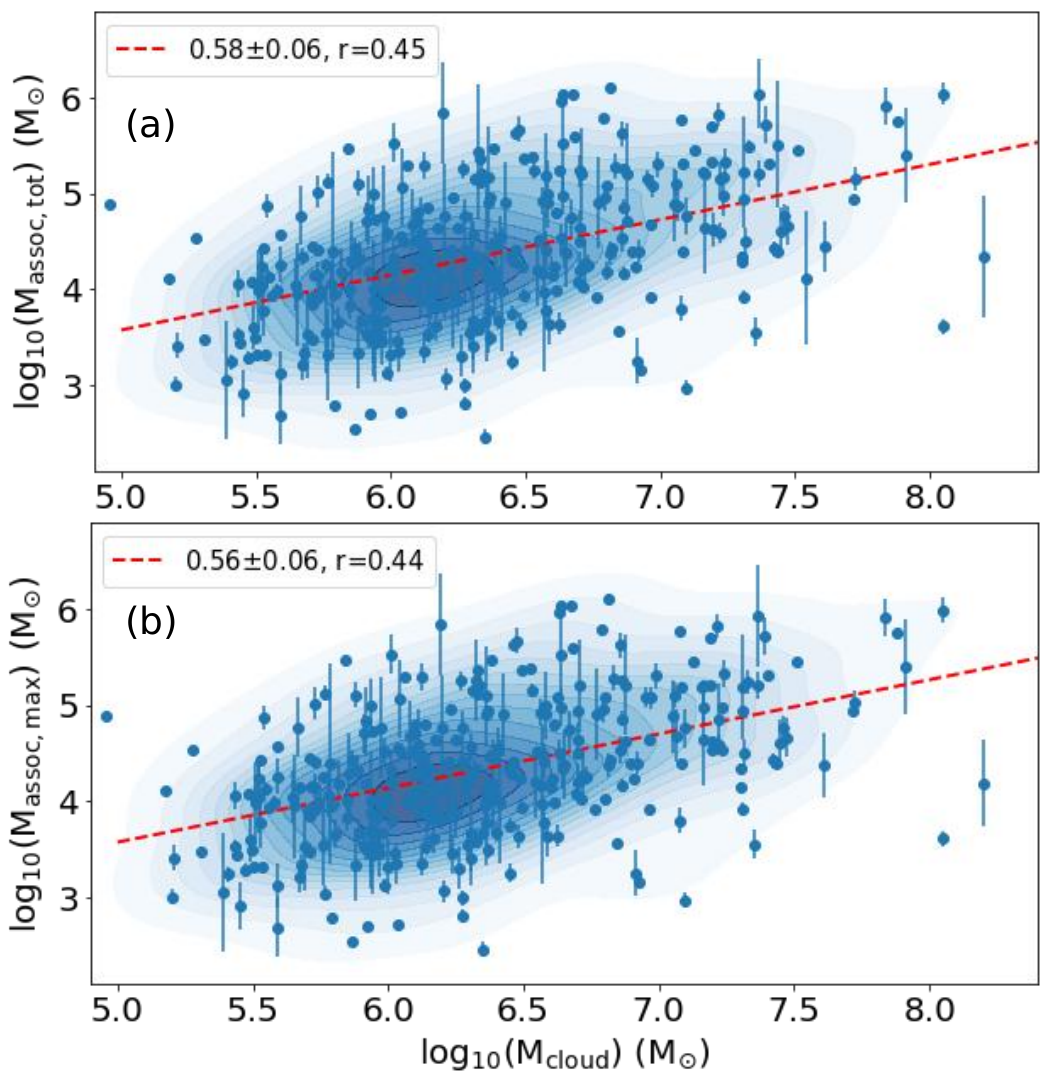}
\caption{(a) The correlation between the mass of the cloud ($M_{\rm cloud}$) and the total mass of the associations in the cloud ($M_{\rm assoc, tot}$), i.e. the $M_{\rm assoc, tot}$-$M_{\rm cloud}$ relation; (b) The correlation between $M_{\rm cloud}$ and the mass of the most massive association ($M_{\rm assoc, max}$) in the cloud, i.e. the $M_{\rm assoc, max}$-$M_{\rm cloud}$ relation. $r$ is the Pearson correlation coefficient.}
\label{assoc}
\end{figure}

Since the beam sizes of all five galaxies are more than 64 pc, we only focus on the associations identified on the scale level of 64 pc. If the distances between the center coordinates of the associations and a molecular cloud are less than the effective radius of the cloud, we distribute these associations to the cloud.
If the associations from the V-band selection and the NUV-band selection overlap, we select the one with a larger area which includes more stars. Actually, the association from the V-band selection always has a larger area, as described in Sec.\ref{catalog}. 
Finally, we added the masses of all independent associations in each cloud. Combining the data of all five galaxies together, Fig.\ref{assoc}(a) shows a clear correlation between the mass of the cloud and the total mass of the associations in the cloud, i.e. the $M_{\rm assoc, tot}$-$M_{\rm cloud}$ relation.
For each cloud,
we also singled out the most massive association with the mass $M_{\rm assoc, max}$. As displayed in Fig.\ref{assoc}(b), the $M_{\rm assoc, max}$-$M_{\rm cloud}$ and $M_{\rm assoc, tot}$-$M_{\rm cloud}$ relations are very similar, and this is because one cloud typically corresponds to one association identified at the 64 pc level.

\subsubsection{Star clusters}\label{s-cluster}

\begin{figure*}
\centering
\includegraphics[width=0.95\textwidth]{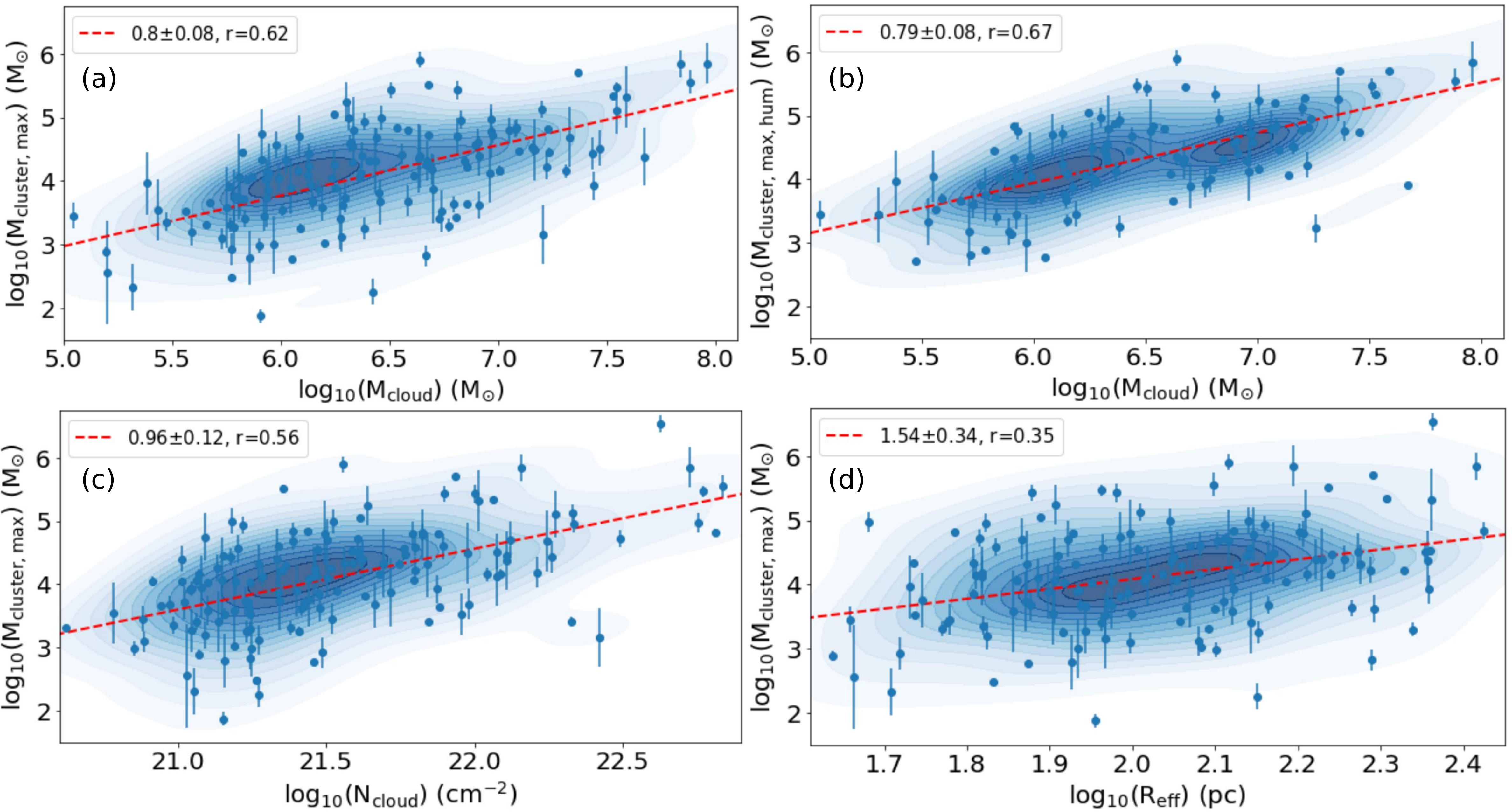}
\caption{(a) and (b) The correlation between the mass of the cloud and the mass of the most massive cluster ($M_{\rm cluster, max}$) in the cloud, i.e. the $M_{\rm cluster, max}$-$M_{\rm cloud}$ relation.
$M_{\rm cluster, max}$ and $M_{\rm cluster, max, hum}$ are derived from the machine learning-classified and the human-classified cluster catalogs
, respectively}; (c) The correlation between the column density of the cloud and $M_{\rm cluster, max}$, i.e. the $M_{\rm cluster, max}$-$N_{\rm cloud}$ relation; (d) The correlation between the effective radius of the cloud and $M_{\rm cluster, max}$. $r$ is the Pearson correlation coefficient.
\label{cluster}
\end{figure*}

Using the same criterion,
if the distances between the center coordinates of the clusters and a molecular cloud are less than the effective radius of the cloud, we distribute these clusters to the cloud. Generally, one cloud will contain multiple star clusters. We first focus on the most massive one. Fig.\ref{cluster}(a) displays a clear correlation between the mass of the cloud and the mass of the most massive cluster in the cloud, i.e. the $M_{\rm cluster, max}$-$M_{\rm cloud}$ relation. 
Moreover, clear correlations are also found between the column density and the size of the cloud ($N_{\rm cloud}$ and $R_{\rm eff}$) and $M_{\rm cluster, max}$, i.e. the $M_{\rm cluster, max}$-$N_{\rm cloud}$ and $M_{\rm cluster, max}$-$R_{\rm eff}$ relations.

\subsubsection{Total stellar populations}\label{s-total}
\begin{figure}
\centering
\includegraphics[width=0.48\textwidth]{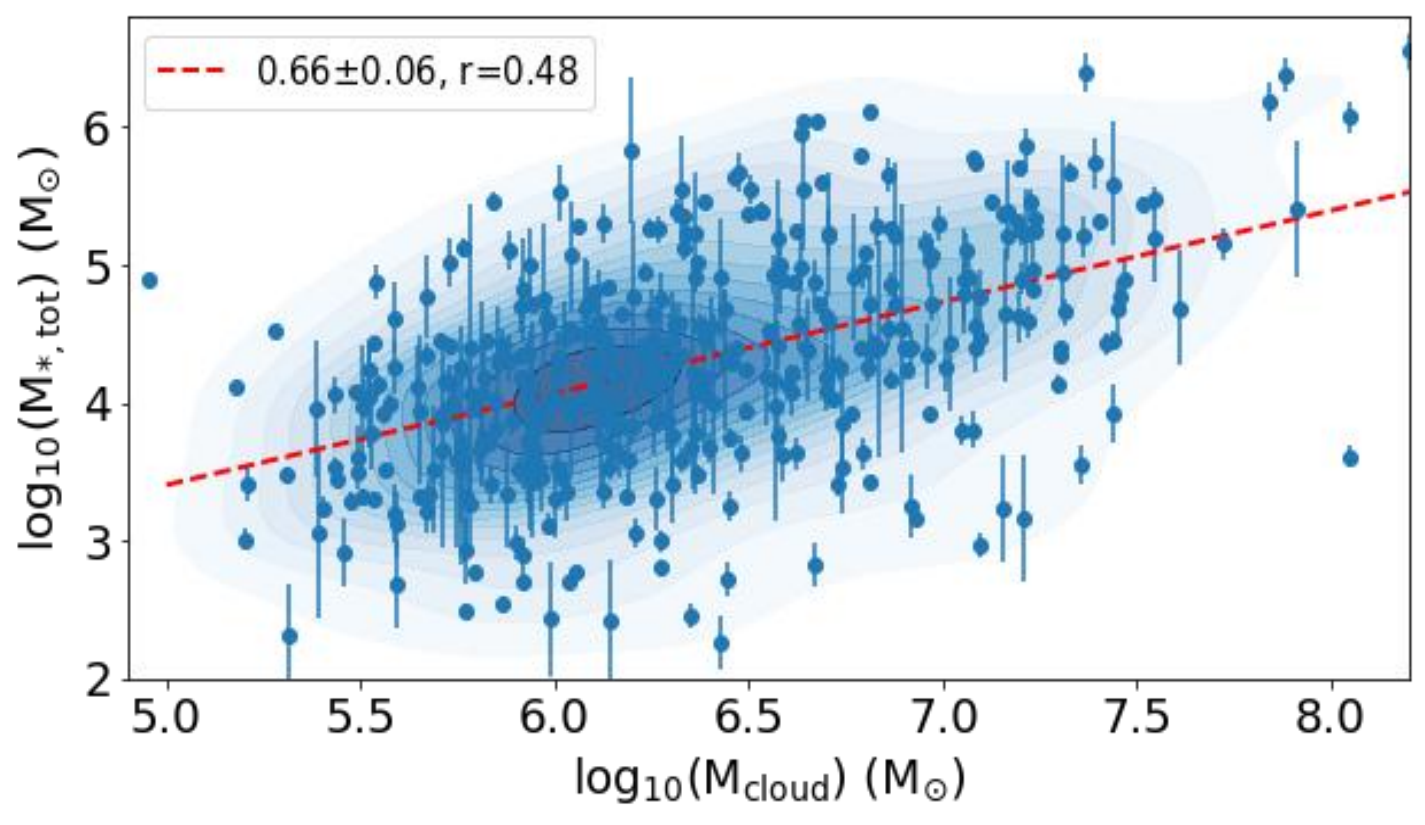}
\caption{The correlation between the mass of the cloud and the mass of the total stellar populations ($M_{\rm *,tot}$) in the cloud, i.e. the $M_{\rm *,tot}$-$M_{\rm cloud}$ relation. $r$ is the Pearson correlation coefficient.}
\label{total}
\end{figure}

A molecular cloud may contain only star clusters (case1), only stellar associations (case2), or both star clusters and stellar associations (case3). For case3, the clusters may be inside the associations or independent of them. To estimate the total mass of the stellar populations within the cloud, we need to sum the masses of both the independent clusters and the associations. To avoid redundant calculations of mass, the clusters that fall within the associations should be excluded. Similarly, if the distances between the center coordinates of the clusters and an association are less than the effective radius of the association, we distribute these clusters to the association and do not count them. 
Fig.\ref{total} displays a clear correlation between the mass of the cloud and the mass of the total stellar populations in the cloud, i.e. the $M_{\rm *,tot}$-$M_{\rm cloud}$ relation. 


\subsection{Embedded stellar populations}\label{s-sfr}

\begin{figure}
\centering
\includegraphics[width=0.48\textwidth]{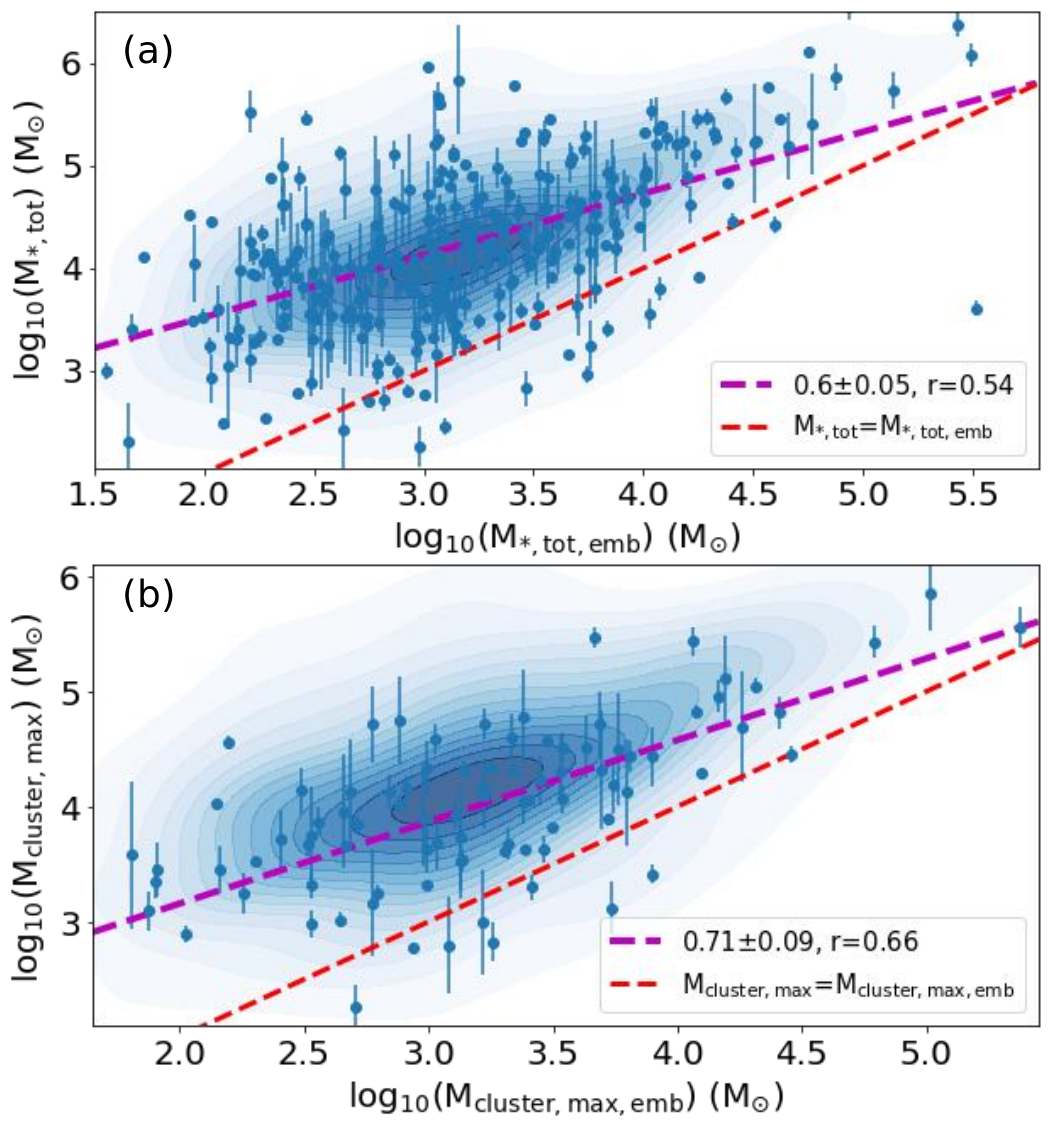}
\caption{(a) Comparison between $M_{\rm *, tot, emb}$ and $M_{\rm *, tot}$; (b) Comparison between $M_{\rm cluster, max, emb}$ and $M_{\rm cluster, max}$. $r$ is the Pearson correlation coefficient.}
\label{emb}
\end{figure}

\begin{figure}
\centering
\includegraphics[width=0.48\textwidth]{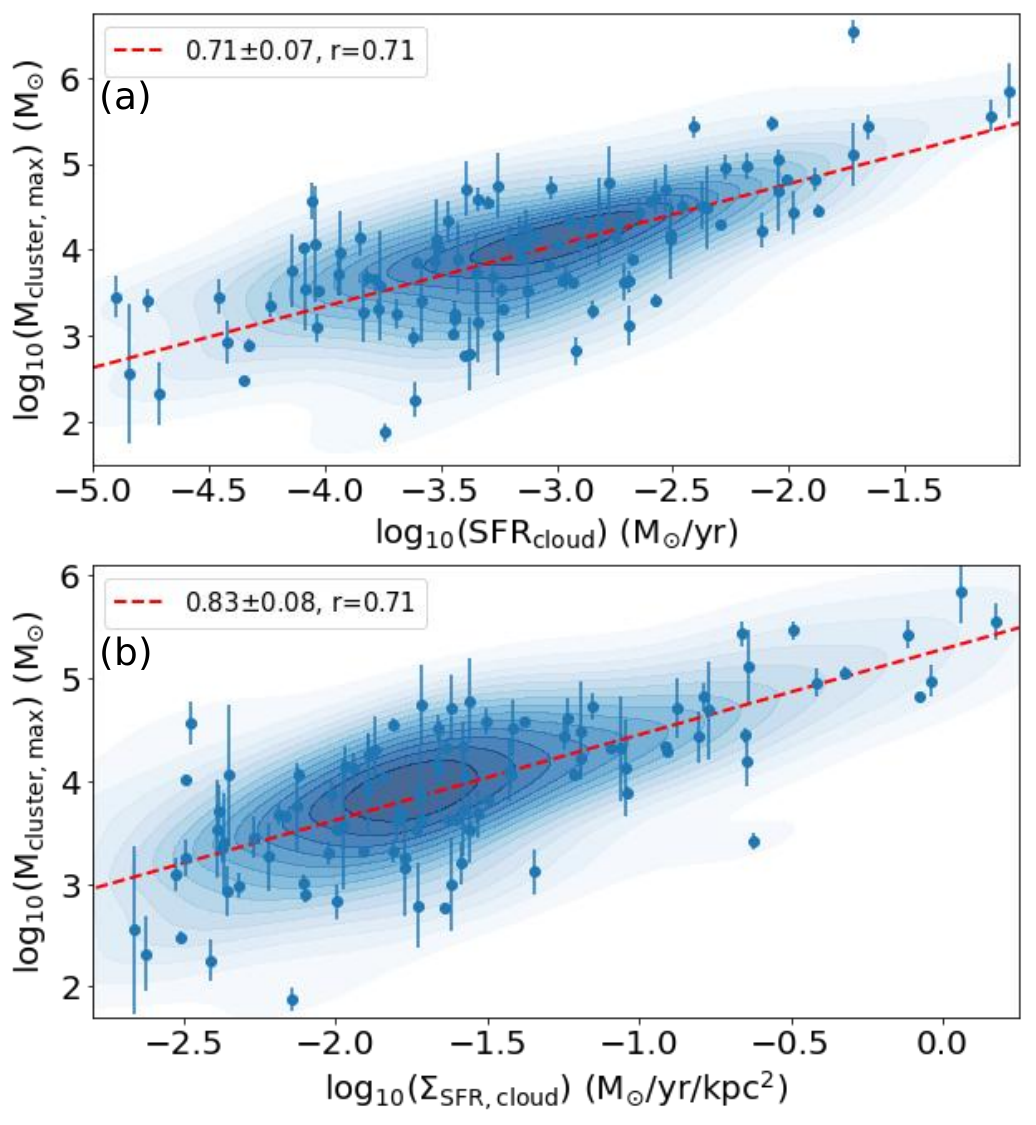}
\caption{(a) The correlation between the star formation rate (SFR) of the cloud and the mass of the most massive cluster in the cloud, i.e. the $M_{\rm cluster, max}$-${\rm SFR}_{\rm cloud}$ relation; (b) The correlation between the SFR surface density of the cloud and $M_{\rm cluster, max}$, i.e. the $M_{\rm cluster, max}$-$\Sigma_{\rm SFR, cloud}$ relation. $r$ is the Pearson correlation coefficient.}
\label{SFR}
\end{figure}

The embedded stellar populations in molecular clouds may not be observed in the PHANGS-HST survey, which is more sensitive to the exposed stellar populations. 
As done in \citet{Zhou2024MNRAS.tmp.2058Z},
the JWST 21 $\mu$m emission was used to estimate the total mass of the embedded stellar populations in a cloud. This method gives quite comparable masses with those obtained from the SED fitting, as compared in \citet{Zhou2024MNRAS.tmp.2058Z}.
Using the prescriptions described in \citet{Leroy2021-257}, the WISE 22 $\mu$m data can be used to calculate the local
star formation rate (SFR) surface density via
\begin{equation}
    \frac{\Sigma_{\rm SFR}}{\rm M_{\odot}~yr^{-1}~kpc^{-2}} = 3.8\times10^{-3} \left(\frac{I_{\rm 22 \mu m}}{\rm MJy~sr^{-1}}\right)\mathrm{cos}i,
    \label{}
\end{equation}
where $i$ is the inclination angle of the galaxy.
Here, we used the JWST 21 $\mu$m data to estimate the SFR surface density.
For NGC 628, \citet{Kim2023-944} defined the duration of the embedded phase of star formation as the time during which CO and 21 $\mu$m emissions are found to be overlapping, i.e. $t_{\mathrm{fb, 21 \mu m}}$, and  they obtained $t_{\mathrm{fb, 21 \mu m}} \approx$ 5.1 Myr. We assume this timescale is also applicable to the five galaxies in this work. Following the same method as in \citet{Zhou2024MNRAS.tmp.2058Z}, the spatial separation between 21 $\mu$m and CO emission peaks is used to estimate the typical age ($t_{\rm age}$) of the embedded stellar populations in the cloud. Then, the total mass of the embedded stellar populations is
\begin{equation}
M_{\rm *, tot, emb}=\Sigma_{\rm SFR} \times A \times t_{\rm age},
    \label{}
\end{equation}
where $A$ is the cloud's area. Here, we consider all 21 $\mu$m emission enclosed in the cloud.
The same method is also applied to calculate the mass of bright 21 $\mu$m sources ($M_{\rm cluster, emb}$) identified in Sec.\ref{dendro}, which represent the embedded clusters. 
As shown in Fig.\ref{emb}, both $M_{\rm *, tot, emb}$ and $M_{\rm cluster, max, emb}$ are systematically smaller than the corresponding $M_{\rm *, tot}$ and $M_{\rm cluster, max}$ observed in the PHANGS-HST survey. This suggests a short duration of the embedded phase.
\citet{Kim2021-504} found that the embedded phase of star formation in six nearby galaxies lasts for 2$-$7 Myr and constitutes 17$-$47 percent of the cloud lifetime. 

Fig.\ref{SFR}(a) and (b) show clear correlations between the SFR and the $\Sigma_{\rm SFR}$ of the cloud and the mass of the most massive cluster in the cloud, i.e. the $M_{\rm cluster, max}$-${\rm SFR}_{\rm cloud}$ relation and the $M_{\rm cluster, max}$-$\Sigma_{\rm SFR, cloud}$ relation.

\subsection{Star formation efficiency}\label{s-sfe}

\begin{figure}
\centering
\includegraphics[width=0.48\textwidth]{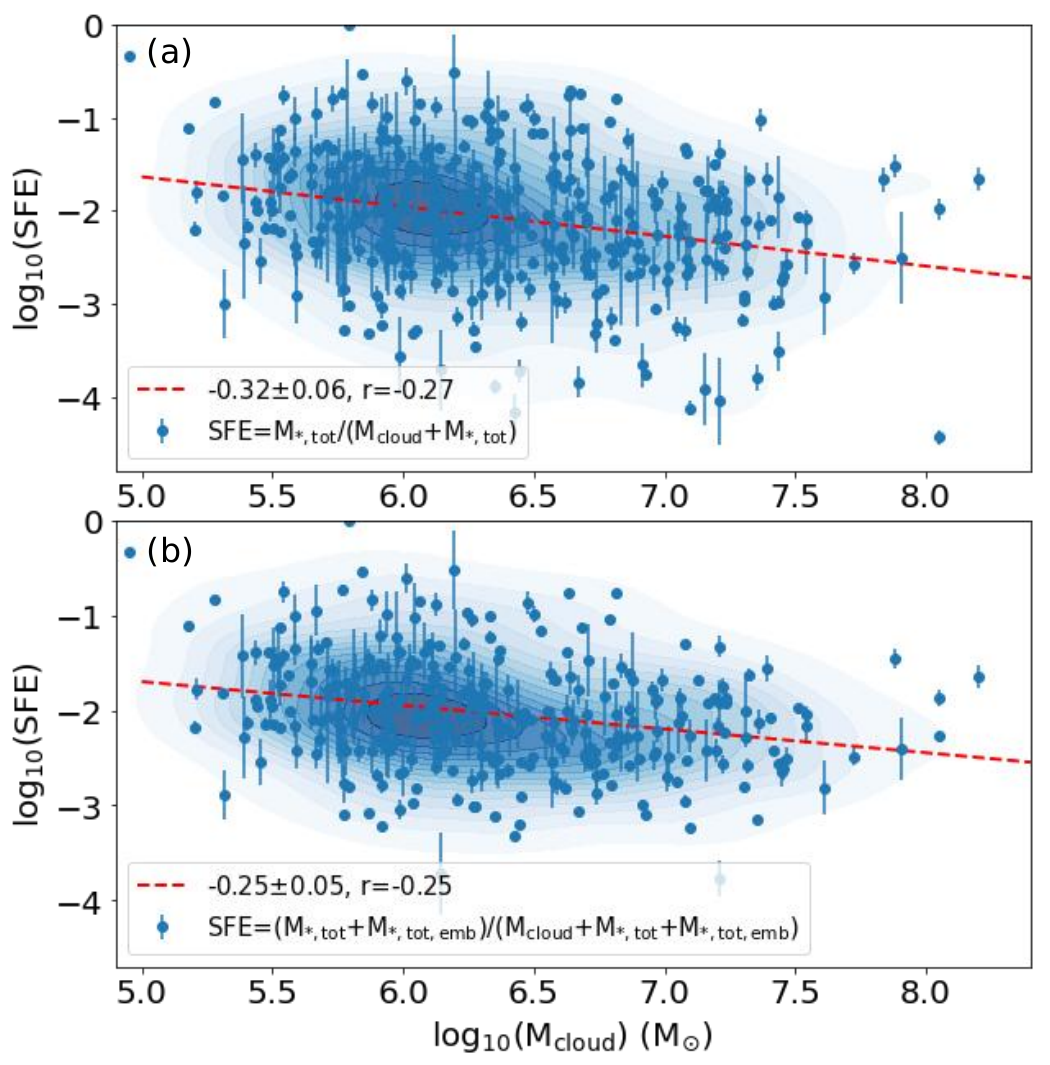}
\caption{The correlation between the star formation efficiency (SFE) of the cloud and the mass of the cloud. (a) Embedded stellar populations were not considered in the calculation of SFE; (b) Embedded stellar populations were considered. $r$ is the Pearson correlation coefficient.}
\label{SFE}
\end{figure}

The star formation efficiency (SFE) of each cloud is defined as
\begin{equation}
{\rm SFE} \approx \frac{M_{\rm *,tot}}{M_{\rm cloud}+M_{\rm *,tot}+M_{\rm I}}, 
\end{equation}
where $M_{\rm I}$ is the mass of the ionized gas. First, we need to assess the significance of $M_{\rm I}$ in the total mass budget.
Based on the HII region catalog in \citet{Santoro2022-658,Groves2023-520}, if the distances between the center coordinates of the HII regions and a molecular cloud is less than the effective radius of the cloud, we distributed these HII regions to the cloud. 
Generally, HII regions associated with molecular clouds are compact. Thus, we used Equation 5 of \citet{Egorov2023-678} to calculate the mass of the ionized gas and this is given by

\begin{equation}
    \label{eq:mass_flux_2}
    M_{\rm I} \approx 29.3 \frac{F{(\rm H\alpha)_{tot}}}{\rm 10^{-16}\ erg\ s^{-1}\ cm^{-2}} \left(\frac{D}{\rm Mpc}\right)^2\left(\frac{n_{\rm e}}{\rm cm^{-3}}\right)^{-1} M_\odot,
    \end{equation}
where $F{(\rm H\alpha)_{tot}}$ is the total $H_{\alpha}$ flux in a HII region. We also use Equation 6 of \citet{Egorov2023-678} to calculate the volume-averaged electron density,
\begin{equation}
\label{eq:dens_hii_vol}
    n_{\rm e} \approx 1.36 \times10^{-16}\left(\frac{L({\rm H\alpha})}{\rm erg\ s^{-1}}\right)^{0.5}\left(\frac{R_{\rm eff}}{\rm pc}\right)^{-1.5} {\rm cm^{-3}}, 
\end{equation}
where $L({\rm H\alpha}) = 4 \pi D^2 F{(\rm H\alpha)_{tot}}$.

Due to the resolution limitations, the radius of the HII region may be overestimated. Additionally, electrons might not be fully distributed throughout the HII region. These factors can lead to an underestimation of the electron density \citep{Egorov2023-678}.
For our samples, the typical value of the calculated electron density is only $\approx$1 cm$^{-3}$. Then we have the typical mass of ionized gas in HII regions being $\approx$10$^{4.5} M_{\odot}$, similar to the value presented in Figure 5 of \cite{Egorov2023-678} calculated by the minimum electron density. Using the maximum electron density, the typical mass of the ionized gas in Figure 5 of \cite{Egorov2023-678} is $\approx$10$^{2.5} M_{\odot}$, which means that the electron density has increased by 2 orders of magnitude overall, i.e. to $\approx$100 cm$^{-3}$. Therefore, when a reasonable estimate of the electron density is available, $M_{\rm I}$ can be neglected relative to $M_{\rm cloud}$. A typical value of $M_{\rm cloud}$ is $\approx$10$^{6.3} M_{\odot}$. Thus, we have
\begin{equation}\label{e-sfe}
{\rm SFE} \approx \frac{M_{\rm *,tot}}{M_{\rm cloud}+M_{\rm *,tot}}.
\end{equation}

Using Eq.\ref{e-sfe}, we estimate the SFE for the clouds in five galaxies. The median value of the SFE in Fig.\ref{SFE} is $\approx$1.4\%, which is comparable with the values listed in \cite{Kim2022-516,Chevance2023-534}. 


\subsection{The ICIMF theory}\label{s-ICIMF}

\begin{table*}
\centering
\small
\caption{All the correlations are fitted with simple linear functions in the form of $y=k*x+b$.  $r$ is the Pearson correlation coefficient.}
\begin{tabular}{ccccccc}
\hline
	$y$ & $x$	&$k$ & $\Delta k$ & $b$ & $\Delta b$ & $r$ \\
log$_{10}$($M_{\rm assoc, tot}$/$M_{\odot}$)&log$_{10}$($M_{\rm cloud}$/$M_{\odot}$)&0.577&0.064&0.685&0.410&0.450\\
log$_{10}$($M_{\rm assoc, max}$/$M_{\odot}$)&log$_{10}$($M_{\rm cloud}$/$M_{\odot}$)&0.563&0.064&0.753&0.411&0.441\\
log$_{10}$($M_{\rm cluster, max}$/$M_{\odot}$)&log$_{10}$($M_{\rm cloud}$/$M_{\odot}$)&0.796&0.082&-1.008&0.536&0.622\\
log$_{10}$($M_{\rm cluster, max}$/$M_{\odot}$)&log$_{10}$($N_{\rm cloud}$/${\rm cm}^{-2}$)&0.961 &0.117 &-16.58 &2.527 &0.561 \\
log$_{10}$($M_{\rm cluster, max}$/$M_{\odot}$)&log$_{10}$($R_{\rm eff}$/${\rm pc}$)&1.539 &0.340 &1.005 &0.693 &0.350 \\
log$_{10}$($M_{\rm *, tot}$/$M_{\odot}$)&log$_{10}$($M_{\rm cloud}$/$M_{\odot}$)&0.663&0.061&0.086&0.395&0.478\\
log$_{10}$($M_{\rm *, tot}$/$M_{\odot}$)&log$_{10}$($M_{\rm *, tot, emb}$/$M_{\odot}$)&0.602&0.054&2.318&0.175&0.536\\
log$_{10}$($M_{\rm cluster, max}$/$M_{\odot}$)&log$_{10}$($M_{\rm cluster, max, emb}$/$M_{\odot}$)&0.709&0.086&1.737&0.285&0.655\\
log$_{10}$($M_{\rm cluster, max}$/$M_{\odot}$)&log$_{10}$[${\rm SFR}_{\rm cloud}$/($M_{\odot} {\rm yr}^{-1}$)]&0.713&0.069&6.192&0.226&0.708\\
log$_{10}$($M_{\rm cluster, max}$/$M_{\odot}$)&log$_{10}$[$\Sigma_{\rm SFR, cloud}$/($M_{\odot} {\rm yr}^{-1} {\rm kpc}^{-2}$)]&0.832&0.081&5.283&0.141&0.709\\
log$_{10}$({\rm SFE})&log$_{10}$($M_{\rm cloud}$/$M_{\odot}$)&-0.320&0.058&-0.034&0.374&-0.267\\
\hline
\label{correlation}
\end{tabular}
\end{table*}

\begin{figure}
\centering
\includegraphics[width=0.48\textwidth]{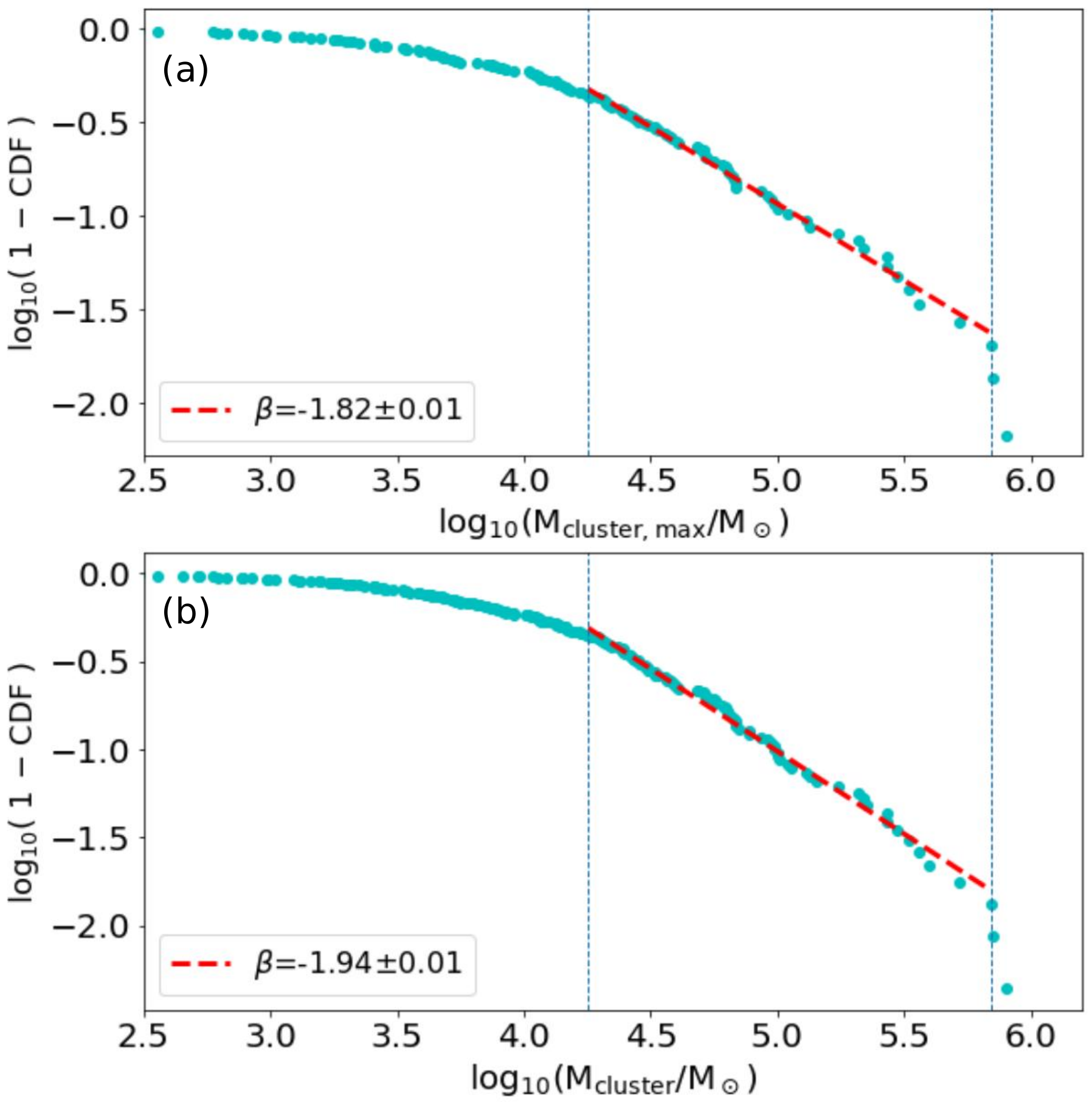}
\caption{
The complementary cumulative distribution function (CCDF) plot and the linear fitting of the cluster mass function.
The vertical dashed lines mark the mass range of the linear fitting.
(a) The mass function of the most massive clusters associated with molecular clouds in Fig.\ref{cluster}(a); (b) The mass function of all clusters associated with molecular clouds.}
\label{cmf}
\end{figure}

\begin{figure}
\centering
\includegraphics[width=0.47\textwidth]{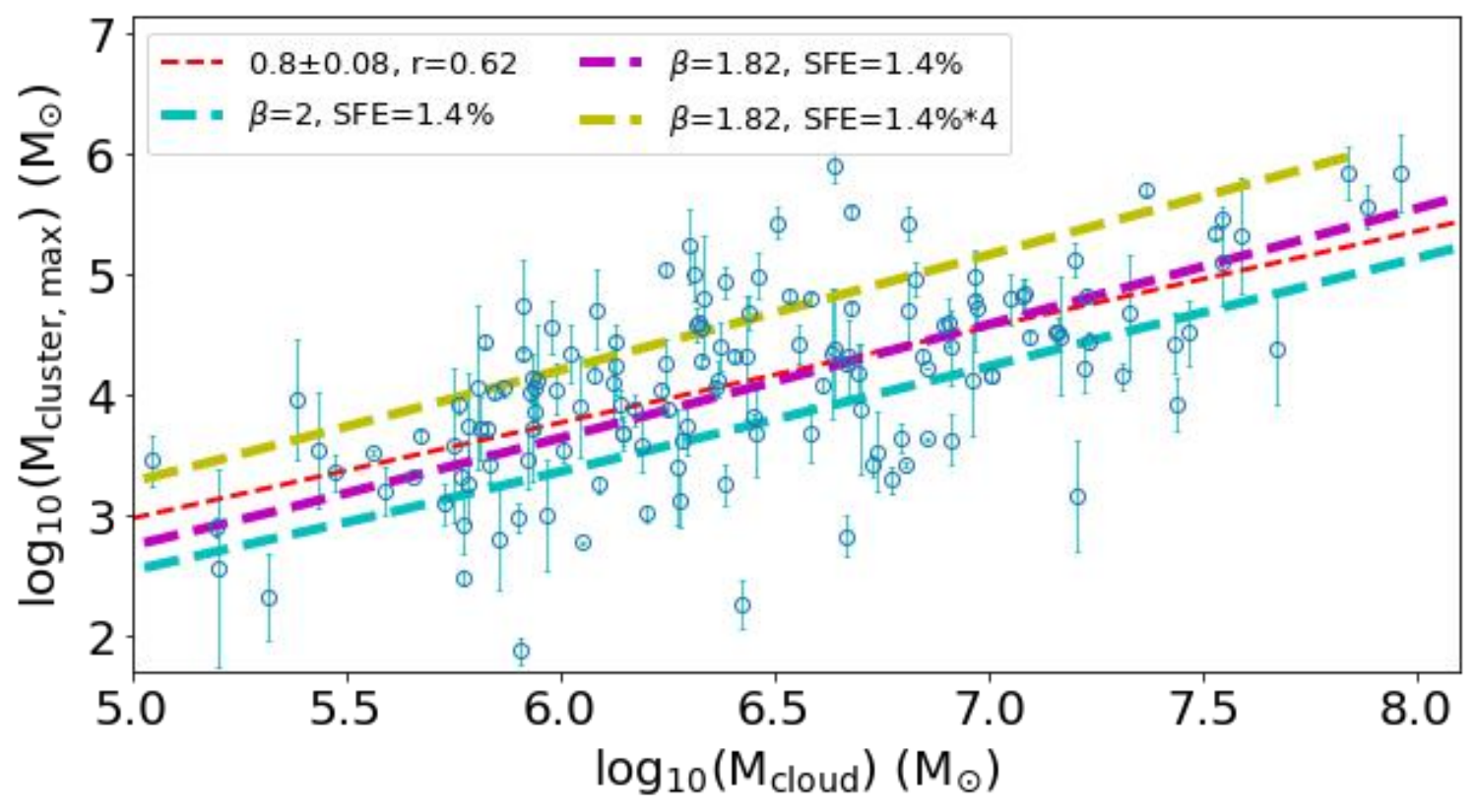}
\caption{The $M_{\rm cluster, max}$-$M_{\rm cloud}$ relation fitted by the model based on the ICIMF theory presented in Sec.\ref{s-ICIMF}. $r$ is the Pearson correlation coefficient. $\beta$ is the slope of the cluster mass function. SFE is the star formation efficiency.}
\label{fit}
\end{figure}


Table.\ref{correlation} summarizes all the fitted correlations between the physical parameters of molecular clouds and their internal stellar populations. 
In this section, we will verify that
the $M_{\rm cluster, max}$-$M_{\rm cloud}$ relation can be understood theoretically on the basis of the integrated
cloud-wide IMF (ICIMF) theory. We start by assuming all the stars in a cloud form in a population of embedded clusters with the mass function
\begin{equation}\label{beta}
\xi_{\mathrm{ecl}}(M_{\mathrm{ecl}})=
k_{\mathrm{ecl}} M_{\mathrm{ecl}}^{-\beta} (M_{\mathrm{ecl,min}} \leqslant M_{\mathrm{ecl}} \leqslant M_{\mathrm{ecl,max}}),
\end{equation}
where $k_{\mathrm{ecl}}$ is a normalization constant and $M_{\mathrm{ecl}}$ is the stellar mass of the embedded cluster.
The mass of the smallest stellar group observed in \cite{Kuhn2015-812} is $\approx 5 M_{\odot}$. Thus, we take $M_{\mathrm{ecl,min}}=5 M_{\odot}$, which is also consistent with the value adopted in \cite{Yan2017-607}.
The embedded cluster population mass conservation gives
\begin{equation}
\label{eq:MtotintMecl}
M_{\mathrm{*,tot}}=\int_{M_{\mathrm{ecl,min}}}^{M_{\mathrm{ecl,max}}}M_{\mathrm{ecl}}~\xi_{\mathrm{ecl}}(M_{\mathrm{ecl}})\,\mathrm{d}M_{\mathrm{ecl}},
\end{equation}
where $M_{\mathrm{*,tot}}$ is the total stellar mass formed in a cloud. 
$M_{\mathrm{*,tot}}$ and the cloud mass are linked by the star formation efficiency, i.e.
\begin{equation}
M_{\mathrm{*,tot}} \approx \mathrm{SFE}*M_{\mathrm{cloud}}.
\end{equation}
To determine the stellar masses within a stellar system based on a given IMF, one needs to apply a specific sampling method. Here, we adopt the optimal sampling. The details of this method and its validity are available in \cite{Kroupa2013-115, Schulz2015-582,Kroupa2015-93,Yan2017-607,Yan2023-670}.
The optimal sampling normalization condition is
\begin{equation}\label{eq:1intMecl}
1=\int_{M_{\mathrm{ecl,max}}}^{M_{\mathrm{ecl,up}}}\xi_{\mathrm{ecl}}(M_{\mathrm{ecl}})\,\mathrm{d}M_{\mathrm{ecl}}.
\end{equation}
We take the upper integration limit (the hypothetical physical upper bound of embedded cluster mass) as $M_{\mathrm{ecl,up}} \approx 10^9$ M$_{\odot}$, following 
\cite{Yan2017-607}. This normalization form is based on the assumption that there exists exactly one object within the mass range from $M_{\mathrm{ecl,max}}$ to $M_{\mathrm{ecl,up}}$ \citep{Schulz2015-582}. 
By solving Eqs.(8)-(11), we can build the relations between $M_{\mathrm{ecl,max}}$ ($\approx M_{\mathrm{cluster,max}}$), $M_{\mathrm{*,tot}}$ and $M_{\mathrm{cloud}}$.
Setting $x \equiv M_{\mathrm{ecl,max}}$, $y \equiv M_{\mathrm{*,tot}}$, $z \equiv M_{\mathrm{ecl,min}}$ and $C \equiv M_{\mathrm{ecl,up}}$, the solution of the equations has two cases:
When $\beta \neq 2$,
\begin{equation}\label{s1}
y= \frac{z^{2-\beta}-x^{2-\beta}}
{\frac{\beta - 2}{1 - \beta}
\left(C^{1 - \beta} - x^{1 - \beta}\right)} \approx
\frac{(\beta-1)(x^{2-\beta}-z^{2-\beta})}{(2-\beta)x^{1-\beta}}.
\end{equation}
When $\beta = 2$,
\begin{equation}\label{s2}
y= \frac{\ln{\left(x \right)}-\ln{\left(z \right)}}
{\frac{1}{1 - \beta}
\left(C^{1 - \beta} - x^{1 - \beta}\right)} \approx x\ln{\left(\frac{x}{z} \right)}.
\end{equation}
In Eq.\ref{s1} and Eq.\ref{s2}, when $\beta>1$,
as long as $M_{\mathrm{ecl,up}}$ is sufficiently larger than $M_{\mathrm{ecl,max}}$, its specific value is not important. 

For the slope $\beta$ of the cluster mass function in Eq.\ref{beta}, generally, the value is $\approx$2 \citep{Lada2003-41,Krumholz2019-57,Mok2020-893,Wainer2022-928}. For massive and compact clusters, such as Super Star Clusters (SSCs), the slope can be $\approx$1.8 \citep{Cuevas2023-525}. Using the most massive clusters associated with molecular clouds presented in Fig.\ref{cluster}(a), we fitted the cluster mass function in Fig.\ref{cmf}(a). It indeed provides a slightly flatter slope, with $\beta \approx$1.82. For all clusters associated with molecular clouds, the slope is $\approx$1.94. However, as shown in Fig.\ref{cmf}, the most massive clusters and all clusters have comparable mass distributions, rather than the most massive clusters having a systematically larger mass. Since we will use the model to fit the $M_{\rm cluster, max}$-$M_{\rm cloud}$ relation, taking the slope fitted in Fig.\ref{cmf}(a) is more appropriate. 
As shown in Fig.\ref{fit}, when fixing SFE=1.4\%, a smaller $\beta$ can provide a better fit. To some extent, a smaller $\beta$ and a larger SFE are equivalent.
For low-mass molecular clouds, larger SFE will yield a better fit. The fact that low-mass molecular clouds have higher SFEs can be observed in Fig.\ref{SFE}.

On the galaxy scale, \cite{Weidner2013-436} suggests the embedded cluster mass function to be flatter (top heavy) for galaxies with a high SFR, with a dependence that is given by
\begin{equation}\label{b-sfr}
\beta=-0.106\log_{10}\mbox{SFR}+2,
\end{equation}
where the SFR is in unit of $[\mathrm{M}_{\odot}/\mathrm{yr}]$. In \cite{Weidner2004-350,Weidner2005-625}, the observed most massive embedded star cluster in a galaxy is determined through the total galaxy-wide SFR
\begin{equation}\label{}
\frac{M_{\rm ecl, max}}{M_{\odot}}\propto\left(\frac{\mbox{SFR}}{\mathrm{M}_{\odot}/\mathrm{yr}}\right)^{0.75}.
\end{equation}
Fig.\ref{SFR}(a) shows a clear correlation between $M_{\rm cluster, max}$ and ${\rm SFR}_{\rm cloud}$, with a slope of $\approx$0.71. Thus, similar to Eq.\ref{b-sfr}, a $\beta$-SFR relation should also exist on the cloud scale.
Surveying adequately complete cluster populations in molecular clouds of nearby galaxies to fit the cloud-scale $\beta$ remains challenging, but it is feasible for molecular clouds in the Milky Way. In order to better fit the observational data in Fig.\ref{fit}, we need to understand how $\beta$ varies with the properties of the molecular clouds. 

Both SFE and
$\beta$ vary with the physical properties of the molecular clouds. Therefore, using the same parameter settings for all molecular clouds can only result in a rough fit. 
Moreover, in this work, we use an unified CO$-$to$-$H$_{2}$ conversion factor (i.e. $\alpha^{2-1}_{\rm CO} \approx 6.7~\rm M_{\odot} \rm pc^{-2} (\rm K~km~s^{-1})^{-1}$) for all five galaxies. 
But $\alpha^{2-1}_{\rm CO}$ may vary with the galaxy, and even different locations within the same galaxy may have different $\alpha^{2-1}_{\rm CO}$. Therefore, it is important to correct $\alpha^{2-1}_{\rm CO}$ for each molecular cloud to accurately estimate the mass of the cloud. This will be left for future work.

\section{Conclusion}\label{conc}

In this study, we first investigated the connections between the physical characteristics of molecular clouds and their stellar populations. We then utilized the ICIMF theory to interpret the observed patterns. 
The main conclusions are as follows:

1. The molecular clouds in five galaxies (NGC 1433, NGC 1566, NGC 1559, NGC 3351 and NGC 3627) were identified based on the integrated intensity (Moment 0) maps of CO (2$-$1) emission from ALMA observations. The catalogs for compact clusters and multi-scale associations across these galaxies were matched with the identified clouds. 
We used the JWST 21 $\mu$m data to estimate the star formation rate (SFR) surface density of the clouds and calculate the masses of the embedded stellar populations in the clouds, and found that the embedded stellar population mass is systematically smaller than that of the exposed stellar population. This suggests a short duration of the embedded phase.
We found that the mass of the most massive cluster ($M_{\rm cluster, max}$) in a cloud is positively proportional to the mass ($M_{\rm cloud}$), the column density, the SFR and the SFR surface density of the cloud. 

2. The mass of the ionized gas ($M_{\rm I}$) in each cloud can be neglected relative to $M_{\rm cloud}$. Based on the masses of the total stellar populations and their corresponding clouds, we obtained a typical value of the cloud-scale SFE, $\approx$1.4\%.  There is a trend that the SFE decreases as the cloud mass increases. More massive clouds lead to a larger stellar population and more massive star clusters, resulting in stronger feedback. This may suppress star formation in massive clouds, leading to a relatively lower SFE. This self-regulated star formation also supports the choice of the optimal sampling.

3. We constructed an integrated cloud-wide IMF (ICIMF) model that provides a good fit to the observed $M_{\rm cluster, max}$-$M_{\rm cloud}$ relation. When fixing SFE$=$1.4\%, there is only one free parameter $\beta$ (the slope of the cluster mass function) in the solutions. For low-mass molecular clouds, a higher SFE can yield a better fit, consistent with the observations.
The ICIMF theory provides a quantitative framework for understanding the correlations between molecular clouds and their internal stellar populations. 


\section{Data availability}

All the data used in this work are available on the PHANGS team website.
\footnote{\url{https://sites.google.com/view/phangs/home}}.

\section*{Acknowledgements}
We would like to thank the referee for the detailed comments and suggestions, which have helped to improve and clarify this work.
We thank Z. Q. Yan for helpful comments.
It is a pleasure to thank the PHANGS team, the data cubes and other data products shared by the team make this work can be carried out smoothly. 

\bibliography{ref}
\bibliographystyle{aasjournal}





\end{document}